\documentclass[fleqn,10pt]{wlscirep}
\usepackage[utf8]{inputenc}
\usepackage[T1]{fontenc}
\usepackage{soul}

\usepackage{bm}

\usepackage{amsmath}

\title{Uncovering anisotropic effects of electric high-moment dipoles on the tunneling current in $\delta$-layer tunnel junctions}

\author[1,*]{Juan P. Mendez}
\author[1,+]{Denis Mamaluy}
\affil[1]{Sandia National Laboratories, Albuquerque, New Mexico, 87123}

\affil[*]{jpmende@sandia.gov}
\affil[+]{mamaluy@sandia.gov}


\begin{abstract}
The precise positioning of dopants in semiconductors using scanning tunneling microscopes has led to the development of planar dopant-based devices, also known as $\delta$ layer-based devices, facilitating the exploration of new concepts in classical and quantum computing. Recently it has been shown that two distinct conductivity regimes (low- and high- bias regimes) exist in $\delta$-layer tunnel junctions due to the presence of quasi-discrete and continuous states in the conduction band of $\delta$-layer systems. Furthermore, discrete charged impurities in the tunnel junction region significantly influence the tunneling rates in $\delta$-layer tunnel junctions. Here we demonstrate that electrical dipoles, i.e. zero-charge defects, present in the tunnel junction region can also significantly alter the tunneling rate, depending, however, on the specific conductivity regime, and orientation and moment of the dipole. In the low-bias regime, with high-resistance tunneling mode, dipoles of nearly all orientations and moments can alter the current, indicating the extreme sensitivity of the tunneling current to the slightest imperfection in the tunnel gap. In the high-bias regime, with low-resistivity, only dipoles with high moments and oriented in the directions perpendicular to the electron tunneling direction can significantly affect the current, thus making this conductivity regime significantly less prone to the influence of dipole defects with low-moments or oriented in the direction parallel to the tunneling.
\end{abstract}

\begin{document}

\flushbottom
\maketitle

\thispagestyle{empty}

\section*{Introduction}\label{sec:introduction}


Atomic precision advanced manufacturing (APAM) has enabled the creation of 2D doped regions, also known as $\delta$-layers, in semiconductors with single-atom precision \cite{Wilson:2006, Warschkow:2016, Fuechsle:2012, Ivie:2021b, Wyrick:2022} and high conductivity \cite{Goh:2006, Weber:2012, McKibbin:2013, Keizer:2015, Skeren:2020, Dwyer:2021}. The APAM is a process to incorporate dopants, such as P or B, at the atomic scale onto Si surface using chemistry surface \cite{Ward:2020,Skeren:2020}. In a simplified way, this process consists in several steps as follows for phosphorus-doped planar structures embedded in silicon (Si: P $\delta$-layer systems) \cite{Wyrick:2019}: We start with a Si surface, normally (100), fully passivated with H; with the tip of a Scanning Tunneling Microscope (STM), it gives the capability to remove H atom by H atom in the exact locations where we want to incorporate the dopants; then, the surface is exposed with a precursor gas, such as phosphene (PH$_3$) \cite{Ward:2020}, followed by a annealing process to incorporate the dopants into the surface; finally, an epitaxial Si is overgrown through a series of annealing processes to protect the planar structure.  

\begin{figure}
  \centering
  \includegraphics[width=0.8\linewidth]{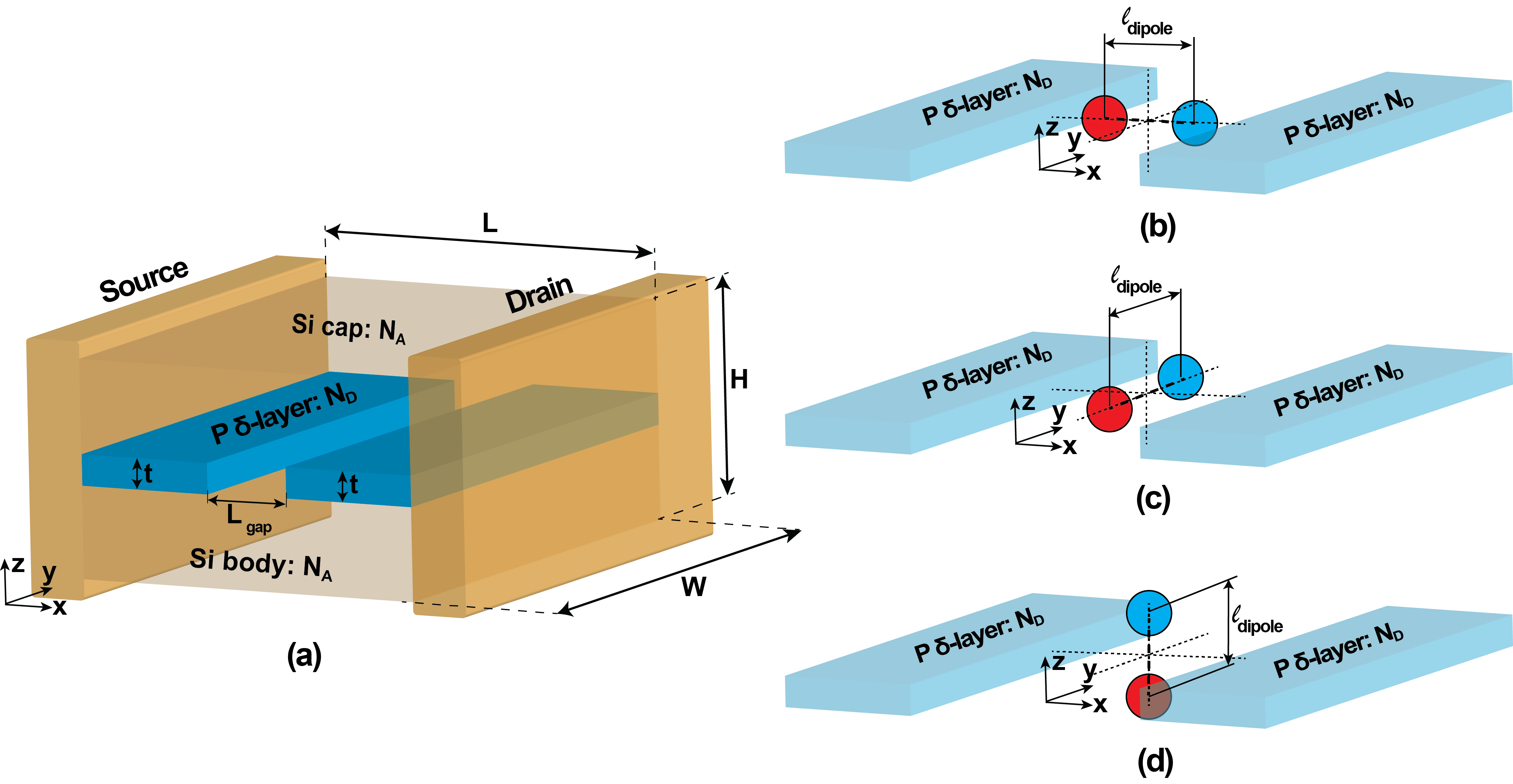}
  \caption{\textbf{Computational model used in our simulations.} A Si: P $\delta$-layer tunnel junction device is shown in \textbf{a}, which consists of a semi-infinite source and drain, in contact with the channel. The channel is composed of a lightly doped Si body and Si cap and two very thin, highly P doped layers with an intrinsic gap of length $L_{gap}$. The representation of an electric dipole oriented along x-direction (propagation direction) is shown in \textbf{b}, y-direction (transverse direction) in \textbf{c}, and z-direction (transverse direction) in \textbf{d}. The negative and positive charged impurities are represented as a blue and red spheres, respectively.}
  \label{fig:TJ_model}
\end{figure}

APAM has various applications, including the exploration of novel electronic devices such as nanoscale diodes or transistors for classical computing and sensing systems \cite{Mahapatra:2011, House:2014, Skeren:2020, Donnelly:2021}. But, most importantly, this technology has been used to explore dopant-based qubits in silicon, with recent advancements in understanding the exchange-based 2-qubit operations \cite{He:2019}, the limits to qubit fidelity from environmental noise \cite{Kranz:2020}, the advantages of leveraging the number of dopants as a degree of freedom \cite{Krauth:2022,Fricke:2021}, and the exploration of many body \cite{Wang:2022} and topological \cite {Kiczynski:2022} effects in dopant chains. One of the building block of these devices is the $\delta$-layer tunnel junction (see Fig.\ref{fig:TJ_model}~\textbf{a}), which consists of two highly doped thin layers separated by an intrinsic gap and embedded in a semiconductor material. These devices require precise control of the tunneling rate for their functioning, as they are very sensitive to tunneling rates. Additionally, it is known that imperfections near the tunnel junction strongly alter the tunneling rate.



\begin{figure}
  \centering
  \includegraphics[width=0.6\linewidth]{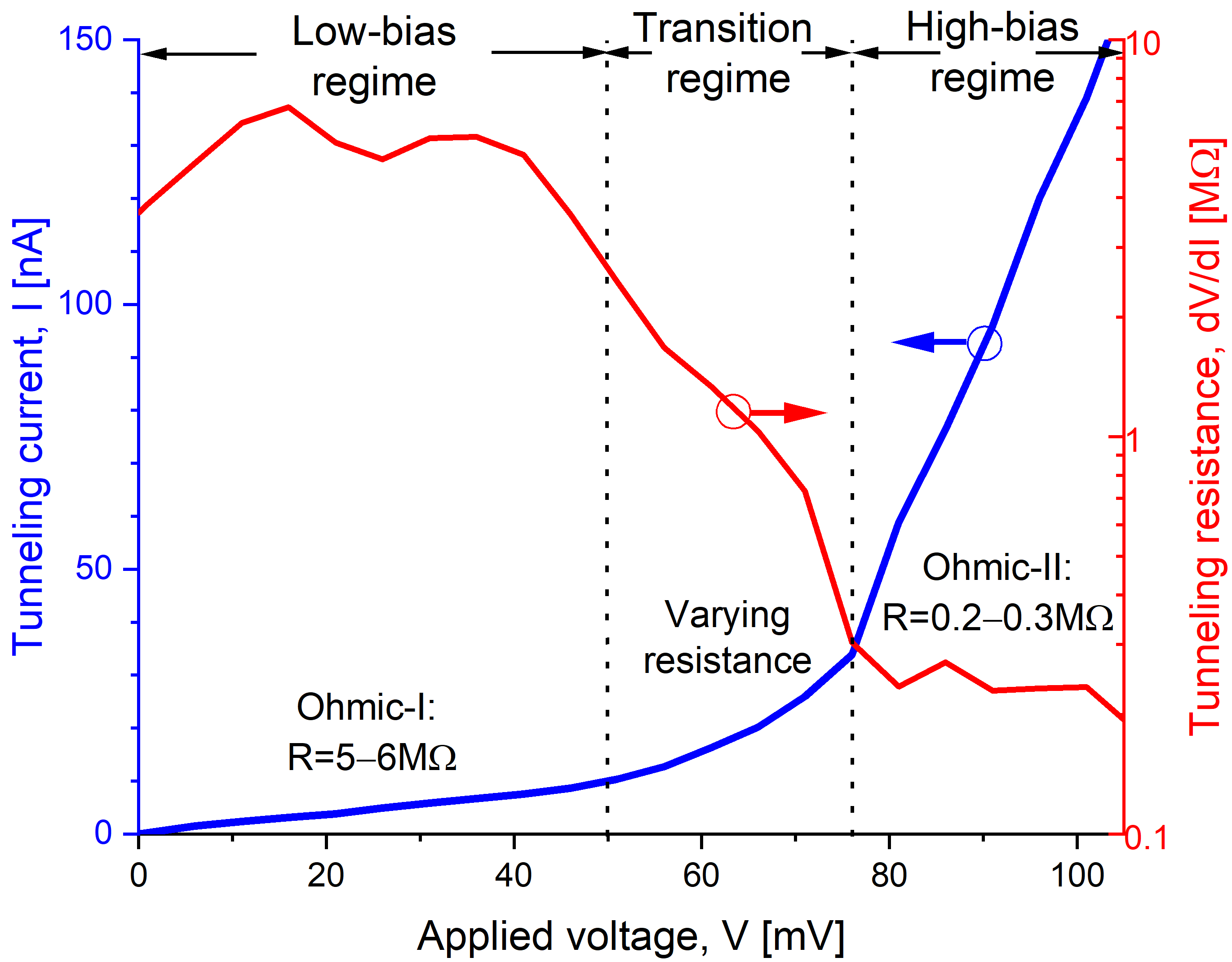}
  \caption{
  \textbf{Characteristic curve for $\delta$-layer tunnel junctions.} Total current $I$ vs. voltage $V$ (blue curve, linear scale) and the corresponding differential resistance $dV/dI$ (red curve, semi-logarithmic scale) are shown for $L_{gap}=10$~nm, $N_D=10^{14}$~cm$^{-2}$, $N_A=5\times 10^{17}$~cm$^{-3}$, $W=15$~nm and $t=1$~nm. Figure reproduced from Ref.~\citen{mendez:2023}. }
  \label{fig:I-V_ideal_TJ}
\end{figure}

In our previous work \cite{Mendez_CS:2022,mendez:2023}, we reported the existence of two different Ohmic regimes in the conductivity of $\delta$-layer tunnel junctions, namely low- and high-bias regimes, connected by a transition regime. For convenience, we include the characteristic IV curve for $\delta$-layer tunnel junctions in Fig.~\ref{fig:I-V_ideal_TJ} reported in Ref.~\citen{mendez:2023}. One can clearly observe these regimes from the result: the first conductivity regime, characterized by a resistance of approximately $5-6$~M$\Omega$, is observed for an applied bias in the range of $0-50$~mV between the source and drain; the second one,  characterized by a resistance of $0.2-0.3$~M$\Omega$, occurs at voltages above $80$~mV; and, these two Ohmic regimes are separated by a transition regime, approximately between $50-80$~mV, where the resistance decreases as the bias is increased. We can understand this phenomenon because of the existence of quasi-discrete states and continuous states in $\delta$-layer systems \cite{Mamaluy:2021,Mendez_CS:2022,mendez:2023}. The threshold voltages of these regimes are determined by the doping density $N_D$ and the $\delta$-layer thickness $t$. Further in Ref.~\citen{mendez:2023}, we investigated how the tunneling rate in $\delta$-layer tunnel junctions can be affected by the presence of charged impurities, specifically focusing on single n-type and p-type impurities in the middle of the tunnel junction. The results, for convenience, are shown in Figs.~\ref{fig:Current_ratio_1mV} and \ref{fig:Current_ratio_100mV} (see continuous and dashed black lines) for distinct tunnel junction lengths. It was found that the tunneling rate is strongly affected by the presence of these charges: the tunneling rate significantly increases by the presence of n-type impurities, while it decreases for p-type impurities.  Additionally, the effect of n-type impurities was found to be stronger than for p-type impurities, specially for large tunnel gaps, in which the effect can be up to one order of magnitude higher. Therefore, these previous works also motivates the present analysis, in which we investigate the influence of electrical dipoles, i.e. zero-charge defects, on the tunneling rate in $\delta$-layer tunnel junction systems. The formation of a dipole occurs when two impurities, one n-type and the other p-type, come into proximity either during the device fabrication process or throughout the device's operational lifespan, as a result of dopant diffusion; Additionally, another possibility is when the formation of a charge-neutral point defect with a non-zero dipole moment. In this work, we focus on the former scenario, and similar conclusions can be drawn for the latter case as well.

In this work, we have employed an efficient implementation of the Non-Equilibrium Green's Function (NEGF) method, referred to as the Contact Block Reduction (CBR) method \cite{Mamaluy:2003,Mamaluy_2004,Mamaluy:2005,Khan:2007,Gao:2014}, to investigate how electric dipoles might alter the tunneling rate in phosphorus $\delta$-layer tunnel junctions in silicon. We have found quantum-mechanical effects, which can not be described by classical methods:  electrical dipoles located near the gap in $\delta$-layer tunnel junctions can significantly alter the current, and its orientation indeed matters. The effect of dipoles oriented in the directions perpendicular to the electron tunneling direction is significantly stronger than for dipoles oriented in the direction parallel to the tunneling direction, revealing an anisotropic effect. This intriguing effect is only present for sufficiently 'large' (a few nanometers or more) dipole moments. For smaller dipole moments, the anisotropic effect vanishes. Similarly, at low bias levels, the anisotropic effect is less pronounced.

\section*{Simulation Setup}\label{sec:simulation setup}

To explore the impact of electric dipoles in the intrinsic gap of a P $\delta$-layer tunnel junction embedded in silicon, we adopt the structure shown in Fig.~\ref{fig:TJ_model}~\textbf{a}. This type of structure is normally referred to as Si: P $\delta$-layer tunnel junction. In the open-system NEGF framework, the computational device consists of a semi-infinite source and drain, in contact with the channel of length $L$, which is composed of a lightly doped Si body and Si cap and two very thin, highly P doped layers (referred to as left and right $\delta$-layers) separated by an intrinsic gap of length $L_{gap}$. The channel length $L$ is chosen to be $30~\text{nm} + L_{gap}$ to avoid  boundary effects between the source and drain contacts, the device height $H$ is $8$~nm, and the device width $W$ is chosen to be $15$~nm, with an effective width of $13$~nm for the $\delta$-layers, to minimize  size quantization effects on the conductive properties\cite{Mendez_CS:2022}. In our analyses, we have considered a thickness of $\delta$-layer of $t=1$~nm, a sheet doping density of $N_D=10^{14}$~cm$^{-2}$ ($N_D^{(2D)} = t \times N_D^{(3D)}$) for the $\delta$-layers, and a doping density of $N_A=5\times 10^{17}$~cm$^{-3}$ in the Si cap and Si body. These doping densities and dimensions are of the order of published experimental work \cite{Ward:2020,Donnelly:2023,Skeren:2020,McKibbin:2009,Polley:2012}. Furthermore, all simulations are carried out at the cryogenic temperature of $4.2$~K, for which we can neglect inelastic scatterings \cite{Goh:2006,Mazzola:2014}.

In this work, we only focus on the effect of dipoles located in the middle of the tunnel gap, with three different orientations, along x-, y- and z-directions, as shown in Fig.~\ref{fig:TJ_model} \textbf{b}, \textbf{c}, and \textbf{d}, respectively, and two distinct electric dipole moments, corresponding to two different scenarios, i.e. with low moment and high moment. The moment of a dipole $p$ is defined as $p=q \times l_{dipole}$, where $q$ is the electric charge and $l_{dipole}$ is the distance between both charges. For this work, we use a dipole length $l_{dipole}$ of 0.8 and $4$~nm, corresponding to the low moment and the high moment, respectively.
When dipoles are oriented along x-direction, we refer to them as dipoles oriented in the propagation direction (i.e. electron tunneling direction), whereas along y- and z-directions we refer to them as dipoles oriented in the transverse directions (i.e. perpendicular to the electron tunneling direction). In these simulations, dipoles are modeled as two point charges of opposing electrical sign, in which each charge is modeled by approximating a point charge with a density of (positive or negative) $4.6 \times 10^{21}$~cm$^{-3}$ homogeneously distributed in a total volume of (0.6~nm)$^3$. The final spatial distribution of the total charge will be dictated by the self-consistent solution of the open-system Schr\"{o}dinger and Poisson equations. Additionally, while our analysis in this work is restricted to the center-gap location, exploring the influence of other locations could be intriguing. This is because free electrons in $\delta$-layer systems form distinct conducting layers perpendicular to the confinement direction, thus signaling a highly non-homogeneous electron density distribution \cite{Mamaluy:2021,Mendez:2020}. 

\section*{Results and Discussion}\label{sec:results and discussion}


\begin{figure}
  \centering
  \includegraphics[width=\linewidth]{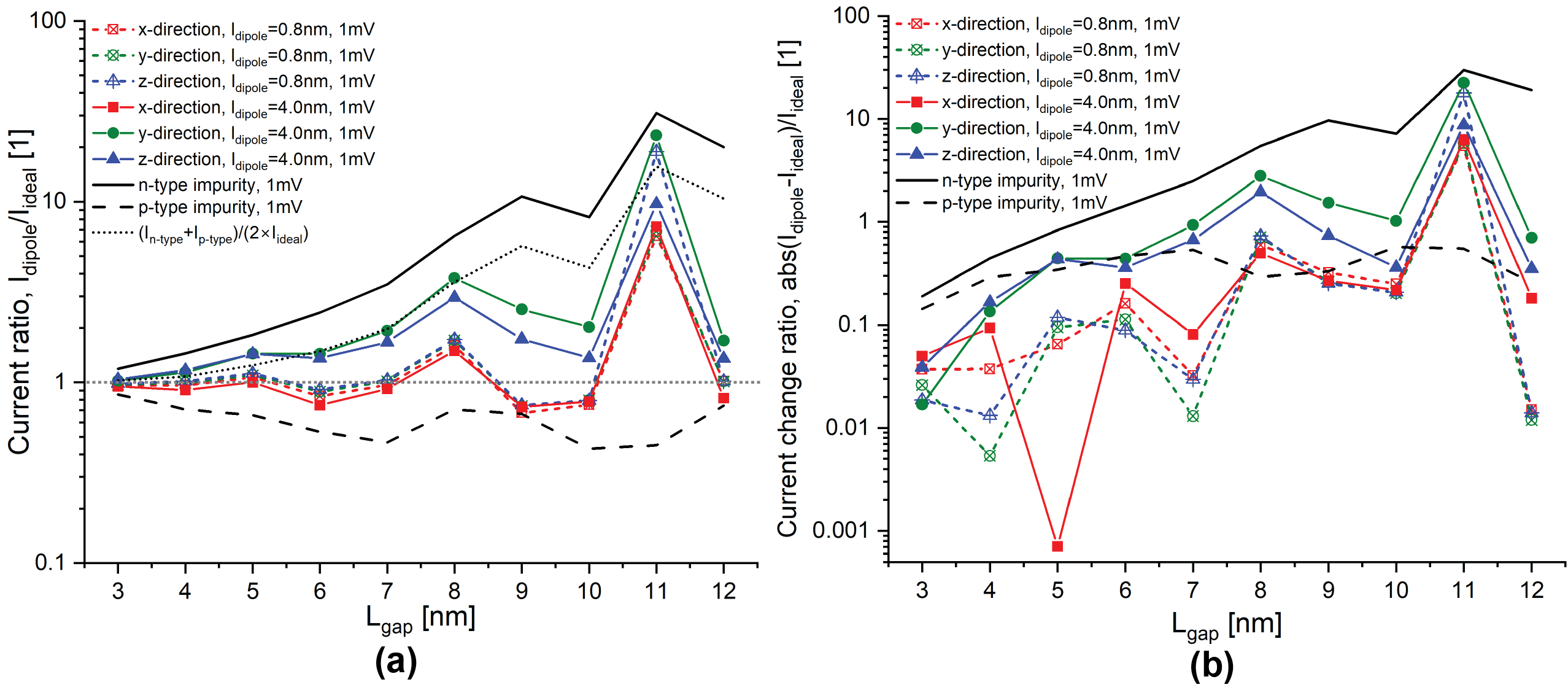}
   \caption{\textbf{Tunneling ratio and tunneling rate change for an applied bias of 1~mV.}Current ratio, $I_{dipole}/I_{ideal}$, between the $\delta$-layer tunnel junction with and without the electric dipole in the intrinsic gap, oriented along the x-, y- and z-directions, for an applied bias of 1~meV in \textbf{a}, and the corresponding tunneling current change in  \textbf{b}. $t=1$~nm, $N_D=10^{14}$~cm$^{-2}$ and $N_A=5\times 10^{17}$~cm$^{-3}$.}
  \label{fig:Current_ratio_1mV}
\end{figure}

\begin{figure}
  \centering
  \includegraphics[width=\linewidth]{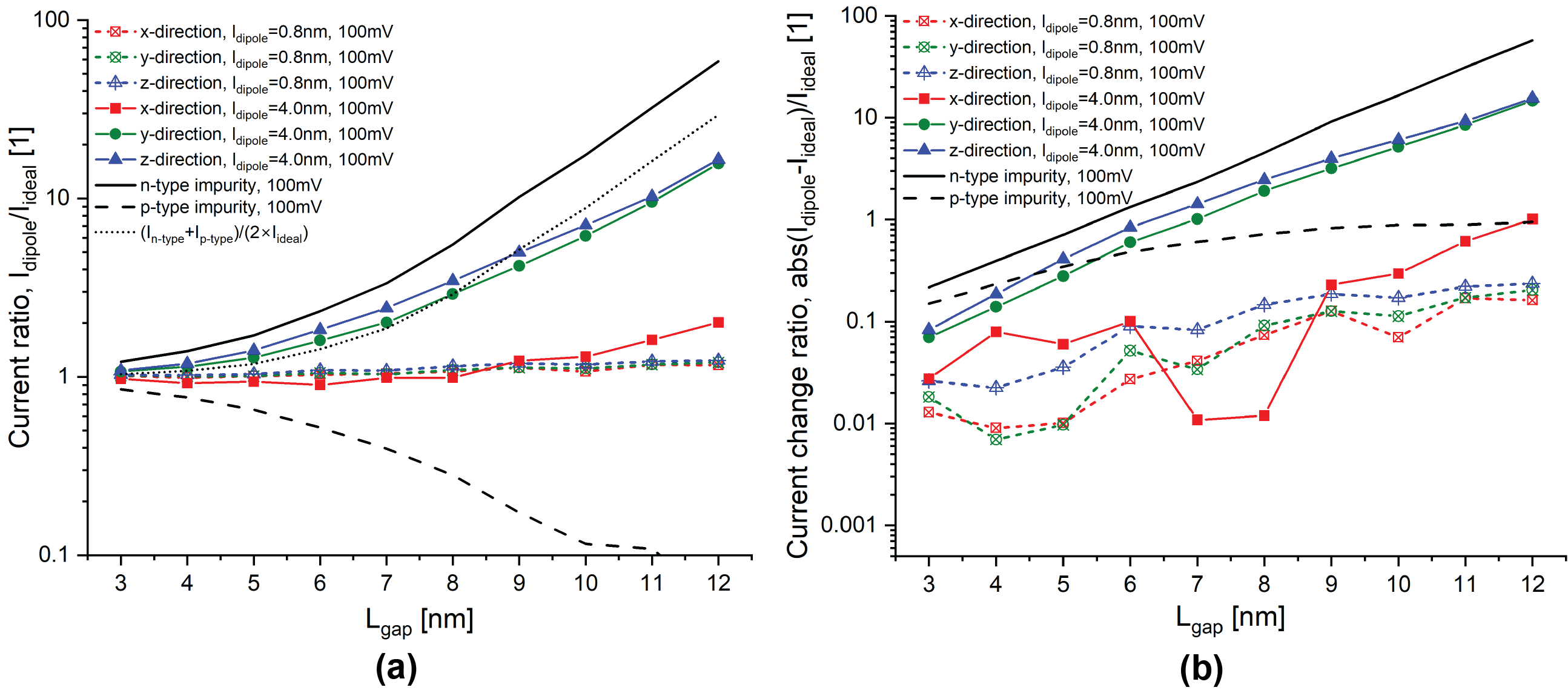}
   \caption{\textbf{Tunneling ratio and tunneling rate change for an applied bias of 100~mV.} Current ratio, $I_{dipole}/I_{ideal}$, between the $\delta$-layer tunnel junction with and without the electric dipole in the intrinsic gap, oriented along the x-, y- and z-directions, for an applied bias of 100~meV in \textbf{a}, and the corresponding tunneling current change in \textbf{b}. $t=1$~nm, $N_D=10^{14}$~cm$^{-2}$ and $N_A=5 \times 10^{17}$~cm$^{-3}$.}
  \label{fig:Current_ratio_100mV}
\end{figure}

We begin by examining the current ratio, defined as the ratio of the current of the system with the dipole to the current of the system without the dipole, and the current change ratio, defined as the ratio of the current change magnitude between the system with and without the dipole to the current of the system without the dipole. Figs.~\ref{fig:Current_ratio_1mV} and \ref{fig:Current_ratio_100mV} show these analyses for an applied bias of $1$ and $100$~mV, respectively, representing the low- and high-bias regimes. Both figures include the analyses for both dipole moments, $l_{dipole}=0.8$~nm for low-moment and $l_{dipole}=4$~nm for high-moment, and all considered orientations, along the propagation direction and along the transverse propagation directions. 
From these results, we can notably discern very interesting results:
i) The trend of the current ratio with the tunnel gap length in the low-bias regime shows a considerable oscillation, and this oscillation vanishes in the high-bias regime;
ii) High-moment dipoles exhibit anisotropic behaviour since dipoles oriented in the transverse propagation directions behave differently than dipoles along the propagation direction; in contrast, low-moment dipoles  does not show that strong anisotropic behaviour since the impact of the dipole orientation is minimum; additionally, high-moment dipoles oriented along the propagation direction behave very similar to low-moment dipoles;
iv) In the high-bias regime, high-moment dipoles oriented in the transverse propagation directions significantly affect the current, whereas the contribution is minimum for all low-moment dipoles and high-moment dipoles oriented along the propagation direction; 
v) In the low-bias regime, specially for large tunnel gap lengths, dipoles of all orientations and moments noticeably affect the tunneling rate in an indistinguishable manner; for narrow tunnel gap length, the effect of all dipoles on tunneling rate is of the same order of magnitude and less pronounced.
In the following, we discuss these intriguing results in more details.

\begin{figure*}
  \centering
  \includegraphics[width=\linewidth]{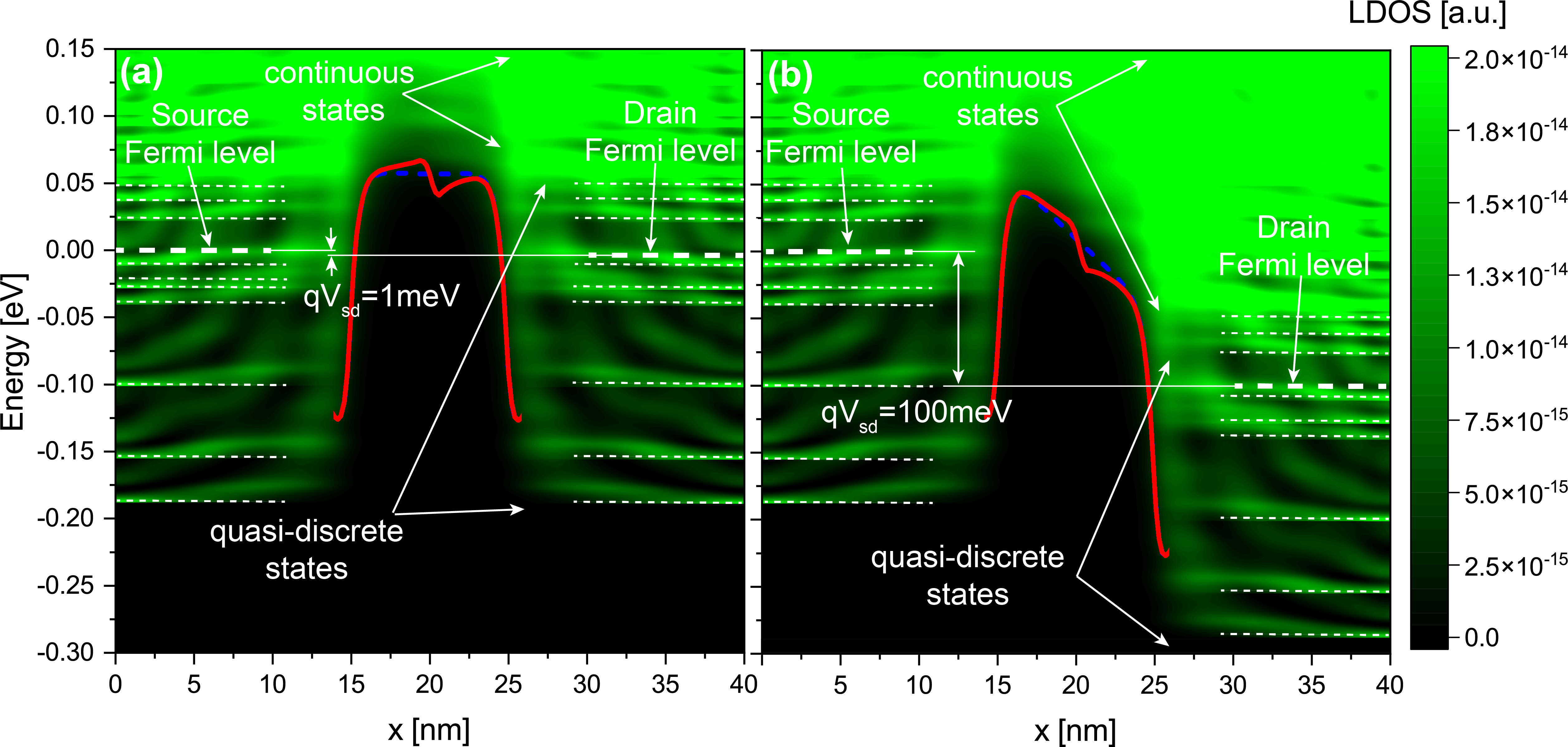} 
  \caption{\textbf{Local Density of States for a $\delta$-layer tunnel junction with a low-moment dipole.} It shows the $LDOS(E,x)$ with a dipole of length $0.8$~nm located  in the middle of the tunnel junction and oriented along x-direction: \textbf{a} and \textbf{b} for an applied voltage of 1 and 100~mV to the drain contact while the source contact is grounded, respectively.  The Fermi levels indicated in the figures correspond to the Fermi levels of the source and drain contacts. The corresponding effective 1D potentials for the ideal device and the device with the dipole are shown as blue dashed and red lines, respectively. $L_{gap}=10$~nm, $N_D=10^{14}$~cm$^{-2}$, $N_A=5 \times 10^{17}$~cm$^{-3}$, and $t=1$~nm.}
  \label{fig:LDOSx_low_moment_dipoles}
\end{figure*}

\begin{figure*}
  \centering
  \includegraphics[width=\linewidth]{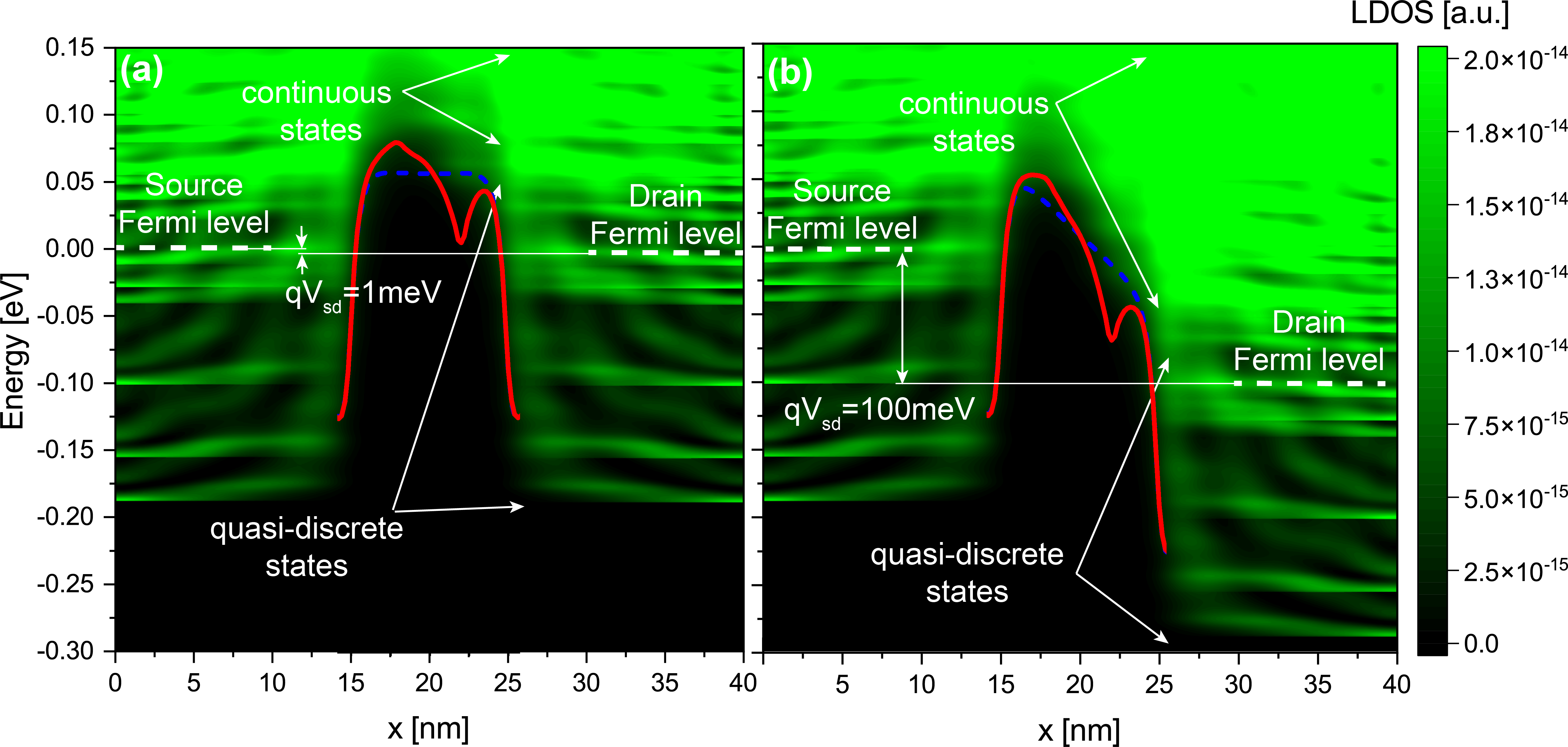} 
  \caption{\textbf{Local Density of States for a $\delta$-layer tunnel junction with a high-moment dipole.} It shows the $LDOS(E,x)$ with a dipole of length $4$~nm located in the middle of the tunnel junction and oriented along x-direction: \textbf{a} and \textbf{b} for an applied voltage of 1 and 100~mV to the drain contact while the source contact is grounded, respectively. The Fermi levels indicated in the figures correspond to the Fermi levels of the source and drain contacts. The corresponding effective 1D potentials for the ideal device and the device with the dipole are shown as blue dashed and red lines, respectively. $L_{gap}=10$~nm, $N_D=10^{14}$~cm$^{-2}$, $N_A=5 \times 10^{17}$~cm$^{-3}$, and $t=1$~nm.}
  \label{fig:LDOSx_high_moment_dipoles}
\end{figure*}

To have a deeper understanding of our analyses, we need to examine the local density of states (LDOS) of the free electrons, which represents the conduction band in real space for the free electrons. Figs.~\ref{fig:LDOSx_low_moment_dipoles} and \ref{fig:LDOSx_high_moment_dipoles} show the $LDOS(x,E)$ along the x-direction for a $\delta$-layer tunnel junction of $L_{gap}=10$~nm with the presence of a dipole of  length $l_{dipole}=0.8,~4$~nm, respectively, in the middle gap: \textbf{a} and \textbf{b} for an applied voltage of 1 and 100~mV to the drain contact while the source is grounded, respectively. In the figures, the states between $x=0-15$~nm and between $x=25-40$~nm correspond to the left and right $\delta$-layers, respectively; the states between $x=15-25$~nm correspond to the intrinsic gap region. At 0~K, the states below Fermi level are occupied, and above the Fermi level are unoccupied. As the temperature increases, some states above the Fermi level start to be occupied in detriment of states below Fermi level. As one can observe from the $LDOS(x,E)$, within the $\delta$-layer regions, the low-energy $LDOS(x,E)$ are strongly quantized, i.e. for energies below the Fermi level and approximately up to $50$~meV above the Fermi level, highlighted with dashed lines in Fig.~\ref{fig:LDOSx_low_moment_dipoles}; for higher energies, approximately above $50$~meV, the $LDOS(x,E)$ are practically continuous in the energy-space dimension, thus signaling that these states are not quantized. Both regions are accordingly marked in the figures. 
The quantization of the conduction band arises from the confinement of dopants in one direction, resulting in very sharp doping profile and leading to the size quantization of the $\delta$-layer \cite{Mendez_CS:2022,mendez:2023}. The presence of discrete states have been observed experimentally in several high resolution Angle-resolved Photoemission Spectroscopy (ARPES) measurements for $\delta$-layers in silicon \cite{Holt:2019,Mazzola:2019}.  Our previous simulations revealed that the number of quantized conduction bands and their corresponding energy splittings are strongly influenced by both the $\delta$-layer thickness $t$ and the doping density $N_{D}$\cite{Mamaluy:2021}. When a positive voltage is applied to the drain contact while the source is grounded, the Fermi level corresponding to the drain contact is shifted down, resulting in lowering the energies of all states in the right side as well. Consequently, new unoccupied states, either in the right $\delta$-layer or those created by the dipole, become available to be occupied by tunneling electrons coming from the left $\delta$-layer.  Thus, we can distinguish two tunneling processes: the first one corresponds to the direct tunneling, i.e. electrons tunneling from occupied states in the left $\delta$-layer to available states in the right $\delta$-layer; and, the second one is the defect-mediated tunneling, i.e. electrons first tunnel from the left $\delta$-layer to the available state in the defect (dipole), and from there to the right $\delta$-layer. Since we are only considering elastic scattering, it is required for both tunneling processes that the energy and momentum are conserved.

The most noticeable observation from Figs.~\ref{fig:Current_ratio_1mV} and \ref{fig:Current_ratio_100mV} is that the current ratio considerably oscillates with the tunnel gap length in the low-bias regime (1~mV), while it becomes very smooth in the high-bias regime (100~mV). This is as a consequence of the strong quantization of the low-energy conduction band in $\delta$-layer systems\cite{mendez:2023}. When a low voltage is applied, e.g. 1~mV, see panel \textbf{a} in Figs. \ref{fig:LDOSx_low_moment_dipoles} and \ref{fig:LDOSx_high_moment_dipoles}, only the unoccupied quantized states in the right $\delta$-layer (those just above the Fermi level), and all quasi-bounded states created by the dipole will play an important role in the tunneling process. If the occupied quasi-discrete states in the left $\delta$-layer or the quasi-bounded states in the dipole align with the unoccupied quasi-discrete states in the right $\delta$-layer, it will result in a considerable increase of the tunneling current. Conversely, if the overlap is minimum, the tunneling current will be reduced. For low biases, this alternating mismatch can only exist for sufficiently large tunnel gaps, because the coupling of the left and right $\delta$-layer wave-functions for narrow tunnel gaps equalizes the electron states on both sides, increasing the overlap and thus eliminating the mismatch. On the other hand, when an high bias is applied, e.g. for $100$~meV, see panel \textbf{b} in Figs.~\ref{fig:LDOSx_low_moment_dipoles} and \ref{fig:LDOSx_high_moment_dipoles}, it makes the continuous unoccupied high-energy states in the right side available for tunneling from the left side, thus diminishing the influence of the conduction band quantization on the current and easing the tunneling process. Thus, this difference in the tunneling mechanism explains the oscillating behaviour of the tunneling ratio in the low-bias regime, specially for large tunnel gaps, and the disappearance of these oscillations in the high-bias regime. Likewise, it also explain the existence of the two conductivity regimes: the first one, between 0 and 50~mV, with high tunneling resistance, where only the quasi-discrete states play a role in the tunneling process; and, the second one, with low tunneling resistance, for voltages above 80~mV, where quasi-discrete and continuous states contribute to the tunneling process.


In the high-bias regime, the effect of low-moment dipoles on the tunneling current is minimum. This finding can be deduced from Fig.~\ref{fig:Current_ratio_100mV}, in which the tunneling ratio for all orientations of low-moment dipole ($l_{dipole}$=0.8~nm) is approximately unity (cf. the results in color dashed lines in \textbf{a}) and the change of the tunneling current is between 1-30\%  (cf. the results in color dashed lines in \textbf{b}) for all studied tunnel junction lengths. Specifically, the tunneling change increases almost exponentially with the tunnel gap length, from a change of 1-3\% for a tunnel gap length of $L_{gap}=3$~nm up to around 20-30\% for  $L_{gap}=12$~nm. In addition, our results also indicate  negligible impact of the dipole orientation on the tunneling since the three studied orientations show similar change in the tunneling rate. These results also indicate that when both impurities are very close to each other, their effects on the tunneling balance out, resulting in minimal changes to the tunneling current, as well as explaining the diminished effect on the tunneling for low-moment dipoles. Furthermore, it is evident from the results that the change in the tunneling current exhibits exponential growth as the tunnel junction length increases. This is primarily due to the dominant effect of the n-type impurity, which exhibits  exponential growth (cf. continuous line in \textbf{b}), while p-type impurity exhibit lower order of growth with the tunnel junction length (cf. dashed line in \textbf{b}).

\begin{figure}
  \centering
  \includegraphics[width=\linewidth]{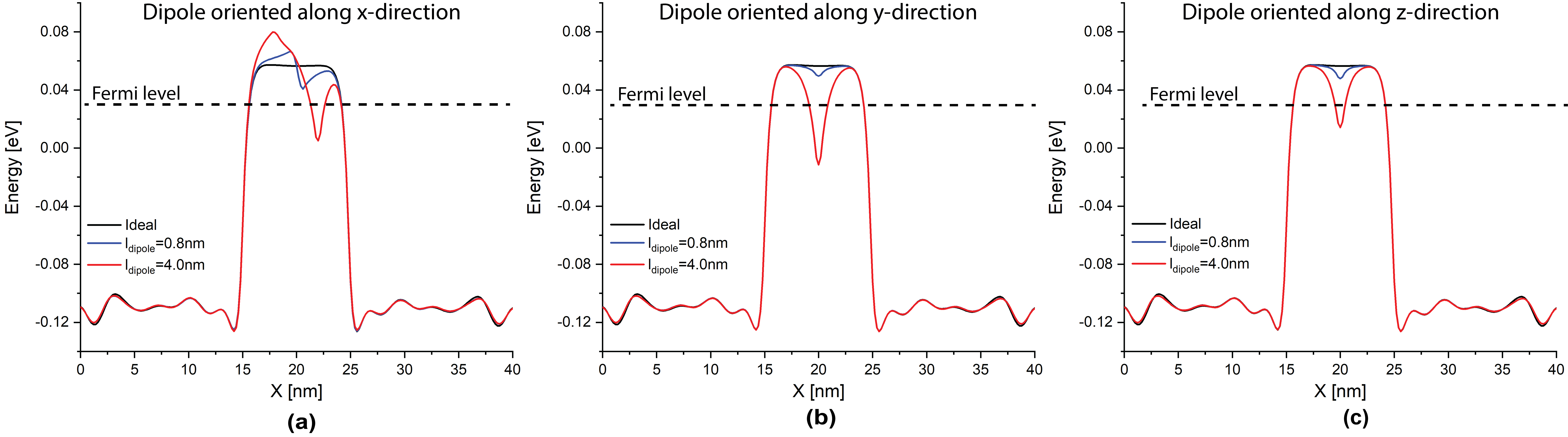}
   \caption{\textbf{Electrostatic potential in equilibrium.} It shows the effective 1D electrostatic potential in equilibrium along the x-direction for a dipole oriented along the x-direction in \textbf{a}, along the y-direction in \textbf{b}, and along the z-direction in \textbf{c}. The results are presented for both dipole lengths, $l_{dipole}=0.8,~4$~nm. The figures also include the electrostatic potential for the ideal $\delta$-layer tunnel junction device represented by the black line. $t=1$~nm, $L_{gap}=10$~nm, $N_D=10^{14}$~cm$^{-2}$ and $N_A=5\times 10^{17}$~cm$^{-3}$.}
  \label{fig:electrostatic potential}
\end{figure}

A different story occurs for high-moment dipoles in the high-bias regime. From Fig.~\ref{fig:Current_ratio_100mV}, we can observe that the orientation of high-moment dipoles in the high-bias regime plays an important role: our results reveals an intriguing anisotropic effect on the tunneling current due to their orientation. When the dipole is oriented along the transverse propagation directions (i.e. y- and z-directions), the tunneling current is strongly affected by the dipole; on contrary, when the dipole is oriented along the propagation direction (x-direction), the tunneling current is only weakly affected. Additionally, it behaves very similar to low-moment dipoles (see red continuous line). In the same manner as low-moment dipoles and for the same reason, the change of the tunneling current increases exponentially with the tunnel gap length. However, for dipoles oriented in the transverse propagation directions, the tunneling change goes from 5-6\% for $L_{gap}=3$~nm up to 2000\% for $L_{gap}=12$~nm, while for dipoles oriented in the propagation direction, the change goes from 3\% for $L_{gap}=3$~nm up to 100\% for $L_{gap}=12$~nm. Furthermore, we can also observe that high-moment dipoles oriented in the transverse propagation directions almost reproduces the dotted line in Fig.~\ref{fig:Current_ratio_100mV} \textbf{a}, which means that the total effect for these dipoles on the current can be approximated as the arithmetic average of independent single n-type and p-type impurities. To understand why the effect of dipoles oriented in the transverse propagation directions is overall stronger than those oriented along the propagation direction, we might need to examine the effective 1D electrostatic potentials in equilibrium, shown in Fig.~\ref{fig:electrostatic potential}. We can see that the barrier height in the electrostatic potential for dipoles in the transverse propagation directions (cf. red lines in panels \textbf{b} and \textbf{c}) is reduced overall, leading to an increase in the tunneling current. Conversely, in the case of dipoles oriented in the propagation direction (cf. red line in panel \textbf{a}), we can see that the barrier height is increased near the p-type impurity and decreased near the n-type impurity. In other words, an increase in the barrier height results in a reduction of the current, while a decrease in the barrier height leads to an increase in the current. Thus, the total tunneling current is a combination of these two opposing effects, resulting in only a minimal change in the tunneling current in this particular case. Comparing the effective 1D electrostatic potentials for low- and high-moment dipoles, we can also note that the peak and dip in the electrostatic potential are less pronounced for low-moment dipoles, revealing again the cancellation of the effects when both impurities are in close proximity to each other. 


Interestingly, in the low-bias regime, specially for larger tunnel junction lengths, dipoles of all orientations and moments affect noticeably the tunneling current in an almost indistinguishable manner. This is because only the quantized conduction band and the quasi-bounded states of the electric dipole are involved on the tunneling process in the low-bias regime (see panel \textbf{a} in Figs.~ \ref{fig:LDOSx_low_moment_dipoles} and \ref{fig:LDOSx_high_moment_dipoles}). Unlike in high-bias regime, the tunneling change is exponential with the tunnel junction length for both dipole moments in a very similar manner. The tunneling change goes from 2-4\% for $L_{gap}=3$~nm up to 620-1800\% for $L_{gap}=11$~nm in low-moment dipoles, and from 2-5\% for $L_{gap}=3$~nm up to 620-2200\% for $L_{gap}=11$~nm in high-moment dipoles. Additionally, it is evident from the results that the anisotropic effect of the orientation of high-moment dipoles is much less significant compared to the high-bias regime.

\begin{figure}
  \centering
  \includegraphics[width=\linewidth]{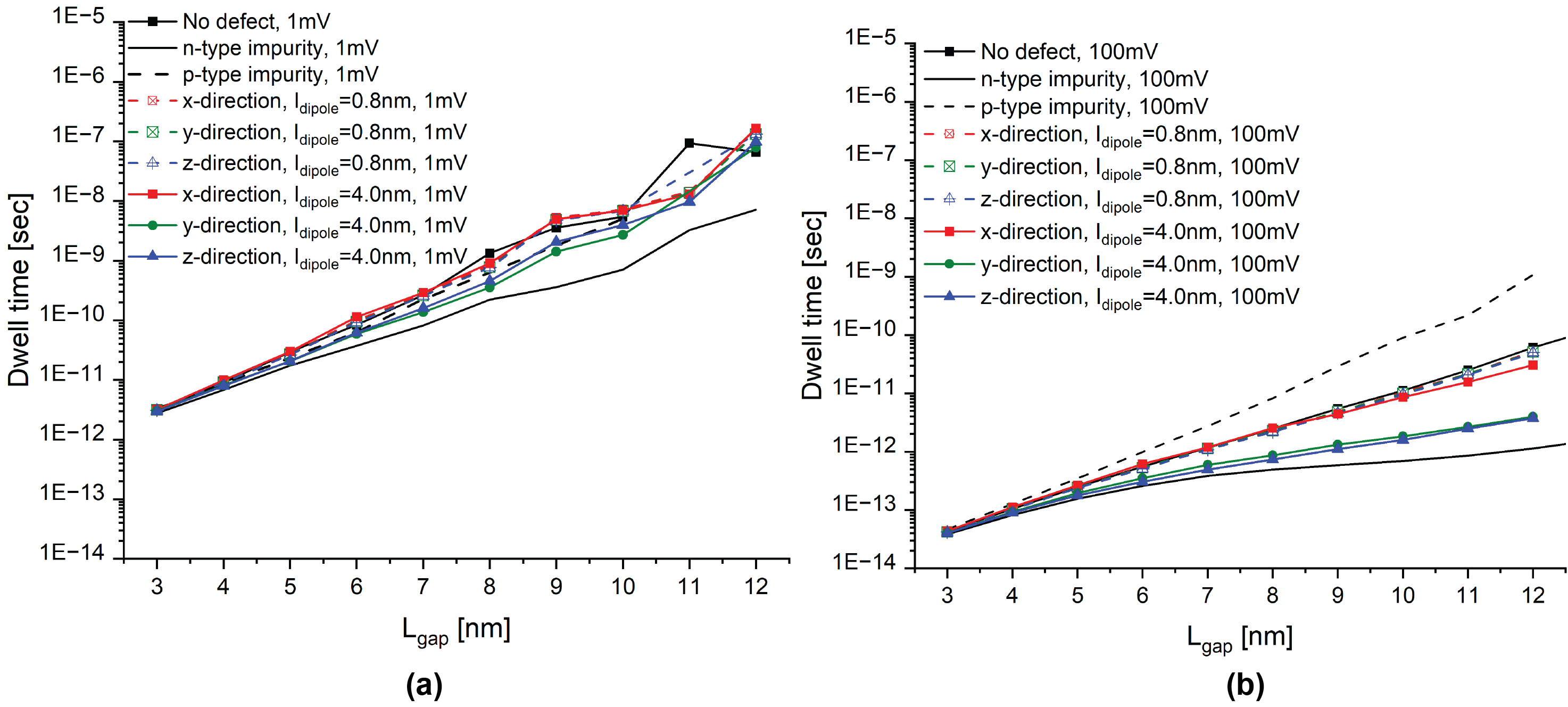}
   \caption{\textbf{Dwell time.} It shows the dwell time of electrons in the tunnel junction in the presence of a dipole in \textbf{a} and \textbf{b} for an applied bias of 1 and 100~mV, respectively. The results are presented for both dipole lengths, $l_{dipole}=0.8,~4$~nm, and all considered orientations. The figures also include the dwell time for the device without defects, as well as with n-type and p-type impurities. $t=1$~nm, $L_{gap}=10$~nm, $N_D=10^{14}$~cm$^{-2}$ and $N_A=5\times 10^{17}$~cm$^{-3}$.}
  \label{fig:dwell time}
\end{figure}

Finally, we investigate the dwell time of the electrons in the tunnel junction, which is the average time that the electrons spend in the barrier, regardless of whether the electrons are reflected or transmitted \cite{dwell_time_Smith,dwell_time_Kelkar_2007,dwell_time_2015}. It is defined mathematically as $t_{dwell~time}=\frac{\int_{\Omega}n(r) \partial \Omega }{I_{tun.~curr.}/q}$, where $I_{tun.~curr.}$ is the tunneling current, $q$ is the elementary charge, $n(r)$ is the electron density and $\Omega$ refers to the tunnel junction domain. From this definition, an alternative physical meaning of the dwell time can be readily deduced, which corresponds to the ratio of the total (average) number of electrons in the junction to the flux (number of electrons per second) going through the system in the steady-state. This ratio is the time necessary to fill the tunnel junction domain $\Omega$ with free electrons to achieve the charge neutrality from the fully depleted state. In other words, it corresponds to the time necessary to switch a device (e.g. FET-type device) from fully "off-" to "on-" state and/or vice versa. Thus, the dwell time can provide insights into the maximum operating frequency of $\delta$-layer tunnel junction devices, which is important for applications. Fig.~\ref{fig:dwell time} shows the dwell time for the free electrons in the tunnel junction in the presence of single charged impurities and single dipole impurities, as well as without defects, in \textbf{a} and \textbf{b} for an applied voltage of 1 and 100~mV, respectively. We can observe several interesting results. Firstly, the dwell time of electrons in the barrier almost decreases inversely proportionally with the applied voltage. This can be explained by the weak increase in the number of electrons within the tunnel junction region with voltage, alongside a corresponding strong (proportional) increase in the current.  Secondly, for the low-bias regime, the dwell time  grows exponentially with the tunnel junction length, and it is almost identical for all cases. However, for the high-bias regime, we observe a strong dependence of the dwell time on the type of the impurity in the tunnel junction. Indeed, Fig.~\ref{fig:dwell time} \textbf{b} shows that a single p-type impurity significantly increases the dwell time, while a n-type impurity significantly decreases the dwell time. The corresponding microscopic interpretation is that a single additional acceptor atom in the gap binds a free electron (with a certain life-time), and thus effectively increases the average time that free electrons spend inside the barrier, whereas a donor has the opposing effect. Similarly, dipole impurities can also significantly affect dwell time of electrons in the barrier. From Fig.~\ref{fig:dwell time} \textbf{b}, it follows that, in the presence of low-moment dipoles and high-moment dipoles oriented along the propagation direction, the dwell times of the electrons are very similar to the dwell time for the ideal device. However, for high-moment dipoles oriented along the transverse propagation directions, the dwell times are noticeably reduced, indicating that electrons spend less time in the barrier.

\section*{Conclusions}\label{sec:conclusions}

In this work we analyze the influence of electric dipole impurities on the tunneling current in $\delta$-layer systems. We employ an efficient implementation of the Non-Equilibrium Green's Function (NEGF) framework, referred to as the Contact Block Reduction (CBR) method \cite{Mamaluy:2003,Mamaluy_2004,Mamaluy:2005,Khan:2007,Gao:2014} to carry out the simulations. Our analysis reveal several interesting results:

i) The trend of the current ratio with the tunnel gap length in the low-bias regime shows a considerable oscillation; this oscillation vanishes in the high-bias regime.

ii) High-moment dipoles exhibit anisotropic behaviour since dipoles oriented in the transverse propagation directions behave differently than dipoles in the propagation direction; in contrast, low-moment dipoles  does not show that strong anisotropic behaviour since the impact of the dipole orientation is minimum; Additionally, high-moment dipoles oriented in the propagation direction behave very similar to low-moment dipoles.

iv) In the high-bias regime, high-moment dipoles oriented in the transverse propagation directions significantly affect current, whereas the contribution is minimum for all low-moment dipoles and high-moment dipoles oriented in the propagation direction. 

v) In the low-bias regime, specially for large tunnel gap lengths, dipoles of all orientations and moments noticeably affect the tunneling rate in an indistinguishable manner; for narrow tunnel gap lengths, the effect of all dipoles on tunneling rate is of the same order of magnitude and less pronounced.

vi) Finally, we also perform the analysis of the dwell times for $\delta$-layer tunnel junctions in the presence of single charged impurities and electrical dipole impurities in the tunnel gap, as well as without defects. Our results indicates the general suitability of $\delta$-layer tunnel junction devices for TeraHertz applications in the "high-conductivity"  regime ($V \geq 80~mV$).

\section*{Method}\label{sec:method}

The simulations in this work are conducted using the open-system charge self-consistent Non-Equilibrium Green's Function (NEGF) Keldysh formalism \cite{Keldysh:1965}, together with the Contact Block Reduction (CBR) method \cite{Mamaluy:2003,Mamaluy_2004,Mamaluy:2005,Khan:2007,Gao:2014,Mendez:2021} and the effective mass theory. The CBR method allows a very efficient calculation of the density matrix, transmission function, etc. of an arbitrarily shaped, multiterminal two- or three-dimensional open device within the NEGF formalism and scales linearly $O(N)$ with the system size $N$. The numerical details of the efficient open-system charge self-consistent treatment in 3D real-space are given in Refs.~\citen{Mamaluy:2021,Mendez_CS:2022,mendez:2023}.

Within this framework, we solve self-consistently the open-system effective mass Schr\"{o}dinger equation and the non-linear Poisson equation \cite{Mamaluy:2003,Mamaluy:2005,Gao:2014}. We employ a single-band effective mass approximation with a valley degeneracy of $d_{val}=6$. For the charge self-consistent solution of the non-linear Poisson equation, we use a combination of the predictor-corrector approach and Anderson mixing scheme \cite{Khan:2007,Gao:2014}. First, the Schr\"{o}dinger equation is solved in a specially defined closed-system basis taking into account the Hartree potential $\phi^H(\bm{r}_i)$ and the exchange and correlation potential $\phi^{XC}(\bm{r}_i)$ \cite{PerdewZunger:1981}. Second, the LDOS of the open system, $\rho(\bm{r}_i,E)$, and the electron density, $n(\bm{r}_i)$, are computed using the CBR method for each iteration. Then the electrostatic potential and the carrier density are used to calculate the residuum $F$ of the Poisson equation
\begin{equation}
\big|\big|\bm{F}[\bm{\phi}^H(\bm{r}_i)]\big|\big|=\big|\big|\bm{A}\bm{\phi}^H(\bm{r}_i) - (\bm{n}(\bm{r}_i)-\bm{N}_D(\bm{r}_i)+\bm{N}_A(\bm{r}_i))\big|\big|,
\end{equation}
where $\bm{A}$ is the matrix derived from the discretization of the Poisson equation, and $\bm{N}_D$ and $\bm{N}_A$ are the total donor and acceptor doping densities arrays, respectively. If the residuum is larger than a predetermined threshold $\epsilon$, the Hartree potential is updated using the predictor-corrector method, together with the Anderson mixing scheme. Using the updated Hartree potential and the corresponding carrier density, the exchange-correlation is computed again for the next step, and an iteration of Schr\"{o}dinger-Poisson is repeated until the convergence is reached with $\big|\big|\bm{F}[\bm{\phi}^H(\bm{r}_i)]\big|\big|<\epsilon=10^{-6}$~eV.
In our simulations we have utilized a 3D real-space model, with a discretization size of 0.2~nm along all directions, thus with about $10^{6}$ real-space grid points, and around 3,000 energy points were used. The CBR algorithm automatically ascertains that out of more than 1,000,000 eigenstates only about 700 ($<0.1\%$) of lowest-energy states is needed, which is generally determined by the material properties (e.g. doping level), but not the device size. We have also employed the standard values of electron effective masses, $m_l = 0.98 \times m_e$, $m_t = 0.19 \times m_e$, the dielectric constant of Silicon, and $\epsilon_{Si}=11.7$. Further details of the methodology can be found in our previous publications \cite{Mamaluy:2021,Mendez:2021,Mendez_CS:2022,mendez:2023}.

\subsection*{Validation}\label{sec:method:validation}

\begin{figure}
  \centering
  \includegraphics[width=0.6\linewidth]{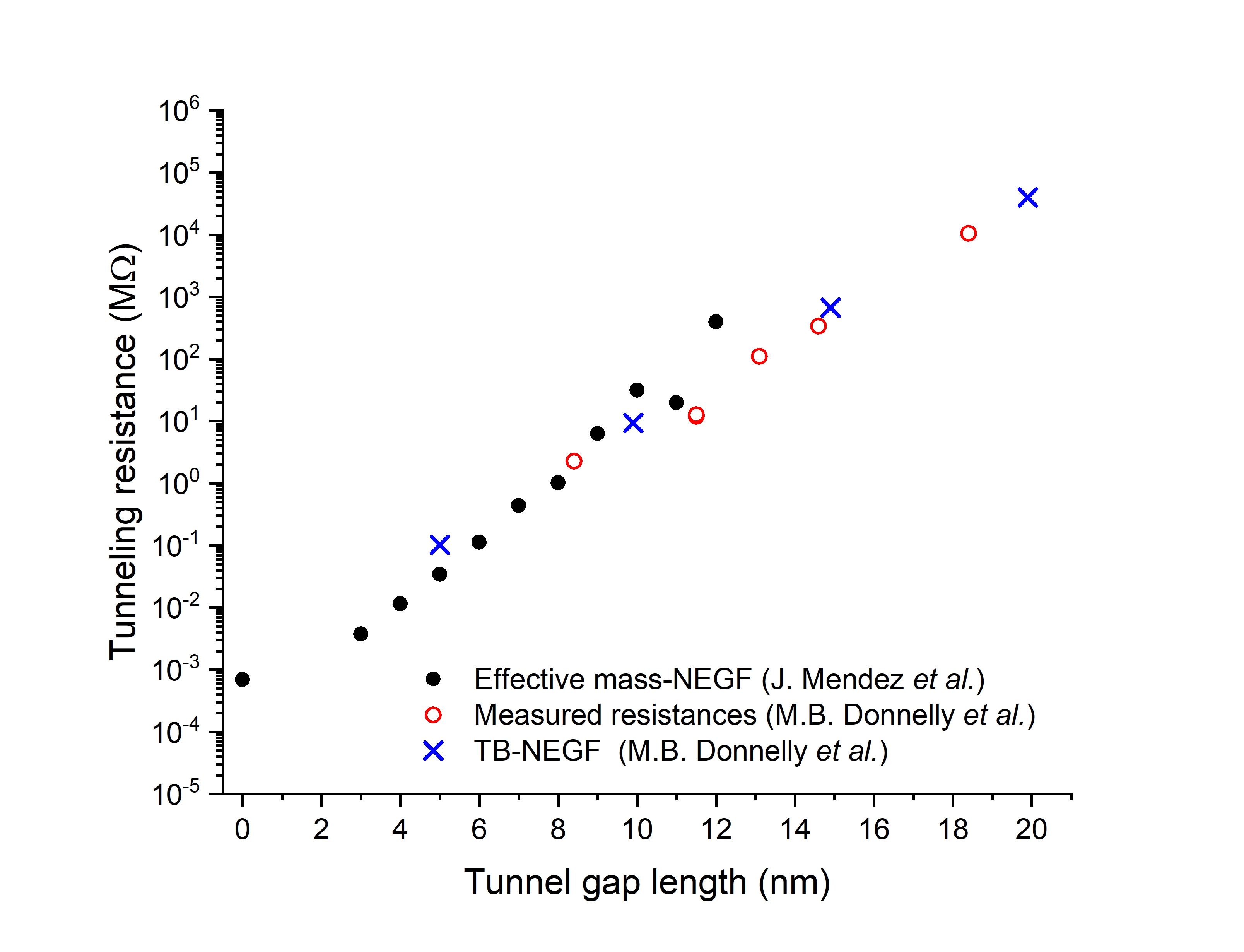}
  \caption{\textbf{Tunneling resistance for Si: P $\delta$-layer tunnel junctions}. Comparison of the resistances computed in our previous work\cite{Mendez_CS:2022}, and measured and computed by M.~Donnelly \emph{et al.}\cite{Donnelly:2023} for diverse tunnel gap lengths. In our study, the dimensions of the $\delta$-layer are $7$~nm in width, $1$~nm in thickness, and a doping density of $N_D=10^{14}$~cm$^2$. In M.~Donnelly \emph{et al.}'s work, the dimensions of the $\delta$-layer are approximately $7$~nm in width, $0.25$~nm in thickness, and a doping density of $N_D=2\times 10^{14}$~cm$^2$.}
  \label{fig:tunneling resistance}
\end{figure}

To validate our computational framework for $\delta$-layer tunnel junctions, we computed the tunneling resistance for different tunnel gaps, $L_{gap}$, and compared the calculations against recently measured data from M.~Donnelly \emph{et al.}\cite{Donnelly:2023}. Fig.~\ref{fig:tunneling resistance} shows our computed tunneling resistance from our previous work\cite{Mendez_CS:2022} for an effective width of the $\delta$-layer of $7$~nm. The figure also includes the resistance measurements and tight-binding calculations for tunnel junctions of $\delta$-layer width of $7$~nm from M.~Donnelly \emph{et al.}\cite{Donnelly:2023}. One can observe that our predicted tunneling resistances (full black circles) are very close to both the experimental data (empty red circles) and the parameter-fitting tight-binding simulations (blue crosses).  Fig.~\ref{fig:tunneling resistance} demonstrates the true predictive power of our open-system effective mass framework for highly-conductive highly-confined systems. We also emphasize that in our calculations there is no fitting parameters, we use only the standard values of electron effective masses and dielectric constant for silicon.  Finally, the slight differences between our computed tunneling resistances and experimental measurements can also be accounted for by the following reasons: i) Certain variations in the width, thickness, and doping density of the $\delta$-layer (note that a doping density of $N_{D}=10^{14}$ was assumed in Ref. \citen{Mendez_CS:2022}, while a higher doping density, $N_{D}=2 \times 10^{14}$, was employed in Ref. \citen{Donnelly:2023}); ii) The possible presence of impurities and/or defects near the tunnel gap.

\section*{Data availability}
The datasets used and/or analysed during the current study available from the corresponding author on reasonable request.

\bibliography{main}

\section*{Acknowledgements}
Sandia National Laboratories is a multimission laboratory managed and operated by National Technology and Engineering Solutions of Sandia, LLC., a wholly owned subsidiary of Honeywell International, Inc., for the U.S. Department of Energy’s National Nuclear Security Administration under contract DE-NA-0003525. This paper describes objective technical results and analysis. Any subjective views or opinions that might be expressed in the paper do not necessarily represent the views of the U.S. Department of Energy or the United States Government.

\section*{Author contributions}
J.P.M. and D.M. equally performed the central calculations and analysis presented in this work.

\section*{Competing interests}
The authors declare no competing interests.

\end{document}